\documentclass[nofootinbib,superscriptaddress]{revtex4}  
\pdfoutput=1
\usepackage{amsmath}
\usepackage{amsfonts}
\usepackage{amssymb}
\usepackage{graphicx, rotating}
\usepackage{epstopdf}
\usepackage{epsfig}
\usepackage{latexsym}
\usepackage{color}
\usepackage{amsmath,amssymb}
\usepackage{slashed}
\usepackage{rotating}
\usepackage[]{hyperref}%
\hypersetup{
colorlinks=true,
linkcolor=red,
anchorcolor=black,
citecolor=blue}
\definecolor{rossos}{cmyk}{0,1,1,0.55}
\definecolor{bluscuro}{rgb}{0.15, 0.2, .85}
\definecolor{bluchiaro}{cmyk}{1,.3,0.,0.1}

\DeclareMathOperator{\GeV}{GeV}
\DeclareMathOperator{\TeV}{TeV}

\newcommand{\bc}{\begin{center}}
\newcommand{\ec}{\end{center}}
\newcommand{\bea}{\begin{eqnarray}}
\newcommand{\eea}{\end{eqnarray}}

\catcode`@=11
\def\marginnote#1{}
\newcommand{\nn}{\nonumber}
\newcount\hour
\newcount\minute
\newtoks\amorpm
\hour=\time\divide\hour by60
\minute=\time{\multiply\hour by60 \global\advance\minute by-\hour}
\edef\standardtime{{\ifnum\hour<12 \global\amorpm={am}%
\else\global\amorpm={pm}\advance\hour by-12 \fi
\ifnum\hour=0 \hour=12 \fi
\number\hour:\ifnum\minute<10 0\fi\number\minute\the\amorpm}}
\edef\militarytime{\number\hour:\ifnum\minute<10 0\fi\number\minute}
\def\draftlabel#1{{\@bsphack\if@filesw {\let\thepage\relax
\xdef\@gtempa{\write\@auxout{\string
\newlabel{#1}{{\@currentlabel}{\thepage}}}}}\@gtempa
\if@nobreak \ifvmode\nobreak\fi\fi\fi\@esphack}
\gdef\@eqnlabel{#1}}
\def\@eqnlabel{}
\def\@vacuum{}
\def\draftmarginnote#1{\marginpar{\raggedright\scriptsize\tt#1}}
\def\draft{\oddsidemargin 0.0truein
\def\@oddfoot{\sl ES, preliminary notes \hfil
\rm\thepage\hfil\sl\today\quad\militarytime}
\let\@evenfoot\@oddfoot \overfullrule 3pt
\let\label=\draftlabel
\let\marginnote=\draftmarginnote
\def\@eqnnum{(\theequation)\rlap{\kern\marginparsep\tt\@eqnlabel}%
\global\let\@eqnlabel\@vacuum} }
\catcode`@=12
\newcommand{\be}{\begin{equation}}
\newcommand{\ee}{\end{equation}}

\newcommand{\eq}[1]{Eq.~(\ref{#1})}

\def\({\left(}
\def\){\right)}
\def\<{\langle}
\def\>{\rangle}
\def\f{\frac}
\def\be{\begin{equation}}
\def\ee{\end{equation}}
\def\bry{\begin{array}}
\def\ery{\end{array}}
\def\bes{\begin{subequations}}
\def\ees{\end{subequations}}
\def\bit{\begin{itemize}}
\def\eit{\end{itemize}}
\def\ben{\begin{enumerate}}
\def\een{\end{enumerate}}
\def\dst{\displaystyle}

\def\Dsl{\slashed{D}}
\def\lra#1{\overset{\text{\scriptsize$\leftrightarrow$}}{#1}}
\def\Xtt{{X_{\hspace{-0.09em}\mbox{\scriptsize2}\hspace{-0.06em}{\raisebox{0.1em}{\tiny\slash}}\hspace{-0.06em}\mbox{\scriptsize3}}}}
\def\Xft{{X_{\hspace{-0.09em}\mbox{\scriptsize5}\hspace{-0.06em}{\raisebox{0.1em}{\tiny\slash}}\hspace{-0.06em}\mbox{\scriptsize3}}}}
\def\Xet{{X_{\hspace{-0.09em}\mbox{\scriptsize8}\hspace{-0.06em}{\raisebox{0.1em}{\tiny\slash}}\hspace{-0.06em}\mbox{\scriptsize3}}}}
\def\Ytt{{Y_{\hspace{-0.09em}\mbox{\scriptsize2}\hspace{-0.06em}{\raisebox{0.1em}{\tiny\slash}}\hspace{-0.06em}\mbox{\scriptsize3}}}}
\def\Yft{{Y_{\hspace{-0.09em}\mbox{\scriptsize5}\hspace{-0.06em}{\raisebox{0.1em}{\tiny\slash}}\hspace{-0.06em}\mbox{\scriptsize3}}}}
\def\Yot{{Y_{\hspace{-0.09em}\mbox{\scriptsize-1}\hspace{-0.06em}{\raisebox{0.1em}{\tiny\slash}}\hspace{-0.06em}\mbox{\scriptsize3}}}}
\def\Ztt{{Z_{\hspace{-0.09em}\mbox{\scriptsize2}\hspace{-0.06em}{\raisebox{0.1em}{\tiny\slash}}\hspace{-0.06em}\mbox{\scriptsize3}}}}
\def\Zft{{Z_{\hspace{-0.09em}\mbox{\scriptsize-4}\hspace{-0.06em}{\raisebox{0.1em}{\tiny\slash}}\hspace{-0.06em}\mbox{\scriptsize3}}}}
\def\Zot{{Z_{\hspace{-0.09em}\mbox{\scriptsize-1}\hspace{-0.06em}{\raisebox{0.1em}{\tiny\slash}}\hspace{-0.06em}\mbox{\scriptsize3}}}}

\def\barXet{{\overline{X}_{\hspace{-0.09em}\mbox{\scriptsize8}\hspace{-0.06em}{\raisebox{0.1em}{\tiny\slash}}\hspace{-0.06em}\mbox{\scriptsize3}}}}
\definecolor{grey}{rgb}{0.6,0.6,0.6}
\definecolor{fuchsia}{rgb}{1,0,1}

\makeatother

\usepackage{amsmath}
\usepackage{amsfonts}
\usepackage{amssymb}
\usepackage{latexsym}
\usepackage{graphicx}
\usepackage{color}
\usepackage{mathrsfs}
\usepackage{slashed}
\usepackage{hyperref}
\usepackage{datetime}
\usepackage{slashed}

\numberwithin{equation}{section}

\hypersetup{colorlinks, citecolor=bluscuro, linkcolor=bluscuro, urlcolor=bluscuro}
\definecolor{rossos}{rgb}{0.8,0.2,0.3}
\definecolor{bluscuro}{rgb}{0.15, 0.2, 0.9}
\definecolor{verdes}{rgb}{0.1, 0.5, 0.1}
\definecolor{myred}{rgb}{0.85, 0, 0}
\definecolor{myblue}{rgb}{0, 0, 0.7}
\definecolor{mygreen}{rgb}{0, 0.45, 0.1}

\makeatother   

 \def\be   {\begin{equation}}   \def\ee   {\end{equation}}
 \def\ba   {\begin{array}}      \def\ea   {\end{array}}
 \def\bea  {\begin{eqnarray}}   \def\eea  {\end{eqnarray}}
 \def\bean {\begin{eqnarray*}}  \def\eean {\end{eqnarray*}}
 \def\nn{\nonumber}

\def\dst{\displaystyle}
\def\f{\frac}

\begin{document}

\title{Composite Charge $8/3$ Resonances at the LHC}
%
\author{Oleksii Matsedonskyi}
\email{oleksii.matsedonskyi@sns.it}
\address{Scuola Normale Superiore, Piazza dei Cavalieri 7, 56126 Pisa, Italy;
Dipartimento di Fisica e Astronomia, Universit\`a di Padova, Via Marzolo 8, I-35131 Padua, Italy}

\author{Francesco Riva}
\email{francesco.riva@epfl.ch}
\address{Institut de Th\'eorie des Ph\'enom\`enes Physiques, EPFL,  CH--1015 Lausanne, Switzerland}

\author{Thibaud Vantalon}
\email{ thibaud.vantalon@epfl.ch}
\address{Institut de Th\'eorie des Ph\'enom\`enes Physiques, EPFL,  CH--1015 Lausanne, Switzerland}

\begin{abstract}
In composite Higgs models with partial compositeness, the small value of the observed Higgs mass implies the existence of light fermionic resonances, the top partners, whose quantum numbers are determined by the symmetry (and symmetry breaking) structure of the theory. Here we study  light top partners with electric charge~$8/3$, which are predicted, for instance, in some of the most natural composite Higgs realizations. 
We recast data from two same sign lepton searches and from searches for microscopic blackholes into a bound on its mass, $M_{8/3} > 940$ GeV. Furthermore, we compare potential reach of these searches with a specifically designed search for three same-sign leptons, both at 8 and 14 TeV. We provide a simplified model, suitable for collider analysis.
\end{abstract}
\keywords{Higgs couplings, Composite Higgs}

\maketitle

\section{Motivation}\label{sec:intro}

A new strongly coupled dynamics which  confines at the TeV scale, inducing the spontaneous breaking of an approximate global symmetry, can produce a naturally light pseudo Nambu-Goldstone boson (pNGB) composite Higgs~\cite{gk,Agashe:2004rs,Contino:2010rs}  at 125 GeV, as required by the experimental observations~\cite{:2012gk}; see  \cite{Bellazzini:2014yua} for a recent review. Although its couplings will deviate slightly from those of the Standard Model (SM) Higgs~\cite{Montull:2013mla,Azatov:2013ura,Barbieri:2013aza},  the best ways to look for this scenario is still  the direct search for other composite resonances generated by the strong sector.

In fact, from an experimental point of view, one of the most important features of a large class of composite Higgs models is the connection, inferable from the symmetry structure, between the smallness of the Higgs mass and the presence of light (lighter than about $1.5$~TeV) fermionic colored composite resonances, interacting predominantly with the third family SM quarks~\cite{lightpartners, Pomarol:2012qf, Redi:2012ha, marzoccaserone, Panico:2012uw}. These are called top partners. Most of the other composite states, on the contrary, are typically expected to have a mass of several TeV, in order to compile with constraints from electroweak  precision observables (see e.g.~\cite{Contino:2013gna}) and flavour physics. This motivates a study of effective field theories including in the spectrum only the lightest layer of composite fermionic partners of the top quark.   

The particular symmetry breaking pattern of these theories, imposed by the necessity of custodial protection in the new, strongly interacting sector, implies that the top-partners would fill multiplets of an approximate symmetry  $\mathcal{H}=SU(2)_L\times SU(2)_R\times U(1)_X$ of which a subgroup $SU(2)_L\times U(1)_Y$ is weakly gauged (the case without custodial symmetry is discussed in Ref.~\cite{Buchkremer:2013bha}). The phenomenology of top partners in the ({\bf 2,2})$_{2/3}$ and ({\bf 1,1})$_{2/3}$ of  $\mathcal{H}$, which are present in most composite Higgs realizations, has already received a lot of attention \cite{Contino:2008hi,DeSimone:2012fs, Vignaroli,azatov} (for the phenomenology of partners of light generations see~\cite{lightgen}), and their signatures are already under experimental scrutiny \cite{lhctoppartners,CMS23, cms53OLD, atlas23,ATLAS:2012hpa};  for studies of  resonances in the ({\bf 3,1}) and ({\bf 1,3}), see Ref.~\cite{Gillioz:2013pba}.

In this article we are interested in models which contain a multiplet of resonances in the ({\bf 3,3})$_{2/3}$, that includes a state with electric charge 8/3. Examples of this are composite Higgs models based on the minimal $SO(5)/SO(4)$ symmetry breaking pattern, in which the top-quark couples linearly to operators in the {\bf 14} of $SO(5)$ \cite{Pomarol:2012qf,Pappadopulo:2013vca,Panico:2012uw}, implying at low energy top-partners 
in the ({\bf 3,3})$_{2/3}$, ({\bf 2,2})$_{2/3}$ and ({\bf 1,1})$_{2/3}$ of $\mathcal{H}$. A similar phenomenology can also appear in models based on the non-minimal cosets, for example $SO(6)/SO(5)$ \cite{Gripaios:2009pe,Frigerio:2012uc} or $SO(6)/SO(4)$ \cite{Mrazek:2011iu}.

The study of the ({\bf 3,3})$_{2/3}$ is in particular motivated by results obtained in a certain class of holographic 5D Composite Higgs models, where this multiplet is found to be the lightest \cite{Pappadopulo:2013vca}. Furthermore, as we will show,  the ({\bf 3,3})$_{2/3}$ is easy to search for, and the bounds that one can potentially extract from these searches are more stringent than those related to other (smaller) multiplets of resonances. For these reasons these searches call for immediate attention.

Stringent constraints on light ({\bf 3,3})$_{2/3}$ arise from electroweak precision observables~\cite{ewpt}, unless other composite resonances compensate their impact. In this article we will assume that such cancellation takes place thanks to the contribution from heavier resonances
that do not affect the LHC phenomenology.

In section~\ref{sec:simplemodel} we propose a simplified scenario with only the charge-8/3 state in the spectrum and just two free parameters. This approximation is particularly suitable for a collider study, and we use it to compare the sensitivity of different searches based on final states with two or three same-sign leptons. We confront this simplified scenario  with the full model in section~\ref{sec:fullmodel} and justify that the former, despite its simplicity, provides a robust model-independent bound on the mass, which is marginally weaker than the one obtained in the more complete set-up.

\section{A Simplified Model}\label{sec:simplemodel}

We propose a simplified model  in which only one colored resonance $\Xet$ with electric charge 8/3 and mass $M_{8/3}$ is present beyond the SM. This model is suitable for a collider analysis and, as shown in the next section, it also captures the most distinctive phenomenological features of complete composite Higgs models (but it also applies to non-composite scenarios~\cite{Bouchart:2008vp}). Due to its large electric charge, no dimension-four interaction can be written between $\Xet$ and the SM field content, and the leading effective interaction must contain
\begin{equation}\label{simpleinteraction}
\frac{c g^2}{\Lambda}\,\barXet \,W_\mu^+\, W^{+\mu}\, t+\textrm{h.c.},
\end{equation}
where $c$ is a dimensionless parameter of order unity, and two powers of the weak coupling $g$ follow naturally from the presence of two gauge bosons. As we will see later, the precise value of $c$ is not important for phenomenology.
The interaction term~(\ref{simpleinteraction}) has to be thought of as originating from an $SU(2)_L$ invariant effective interaction of a more complete model, such as for instance  the dimension-5 coupling of \eq{dd83}, or the one of \eq{5383}, where the intermediate $\Xft$ has mass $M_{5/3}=\Lambda\gg M_{8/3}$ and has been integrated away (in the next section we will also comment on the case where $M_{5/3}< M_{8/3}$). 
In  scenarios where only the multiplet containing the $\Xet$ is present in the low energy spectrum, as we consider in this work, the top quark in the Eq.~\ref{simpleinteraction} is right-handed. 
This assumption, which we will adopt also for this simplified model, has a non-negligible impact on the phenomenology. 
For instance, as was shown for models with charge-5/3 states coupled to the right-handed top, a signal acceptance in the two same sign leptons channel is approximately 10-20\% higher than for left-handed tops~\cite{DeSimone:2012fs}; we expect to have a similar difference in the $\Xet$ case.

At the LHC, the $\Xet$ resonances would be principally produced in pairs via QCD interactions: like all top partners, the $\Xet$ has the same $SU(3)_C$ quantum numbers of top quarks and their parton-level pair production cross section depends uniquely on their mass (NNLO pair production cross section for the 8 and 14 TeV LHC are summarized in table~\ref{hator}). In fact, the $\Xet$ single production with $W^+ t\to \Xet  W^-$ or  $W^+W^+\to \Xet \bar t$ topologies, is suppressed with respect to  pair production by the scale $\Lambda$ and by an additional power of the weak coupling. For this reason we will neglect its effect in what follows.
\begin{table}[ht]
\begin{center}
\begin{tabular}{|c||c|c|}
\hline
$M$&8 TeV & 14 TeV \\
\hline
\hline
600&168.7&1459\\
\hline
700&56.40&581.4\\
\hline
800&20.53&254.3\\
\hline
900&7.943&119.4\\
\hline
1000&3.213&59.21\\
\hline
1100&1.341&30.68\\
\hline
1200&0.573&16.47\\
\hline
\end{tabular}
\hspace{1cm}
\begin{tabular}{|c||c|c|}
\hline
$M$&8 TeV & 14 TeV \\
\hline
\hline
1300&0.248&9.101\\
\hline
1400&0.108&5.149\\
\hline
1500&0.047&2.971\\
\hline
1600&0.020&1.743\\
\hline
1700&0.009&1.036\\
\hline
1800&0.004&0.623\\
\hline
1900&0.001&0.378\\
\hline
\end{tabular}
\caption{\emph{NNLO pair production cross-section $\sigma(M)$ in $fb$ for colored fermions of  mass $M$, calculated using the HATHOR~code~\cite{hathor}, using MSTW2008 parton distribution functions (PDF) \cite{MSTW}.}}\label{hator}
\end{center}
\end{table}%

\begin{figure}[!t]
      \includegraphics[width=0.35\textwidth]{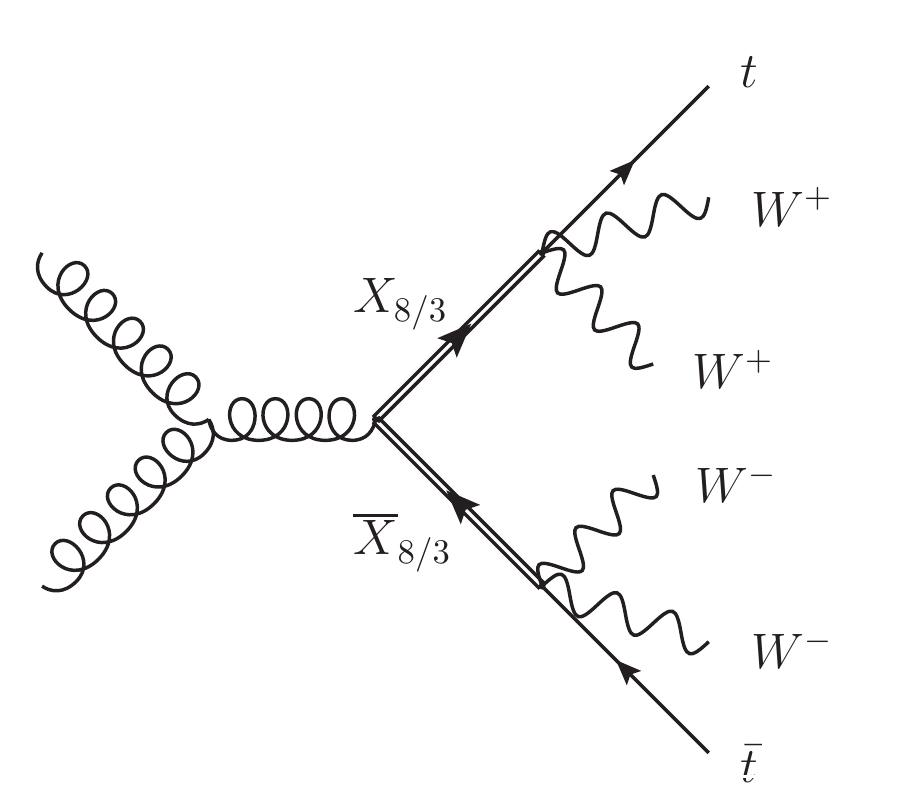}
       \includegraphics[width=0.35\textwidth]{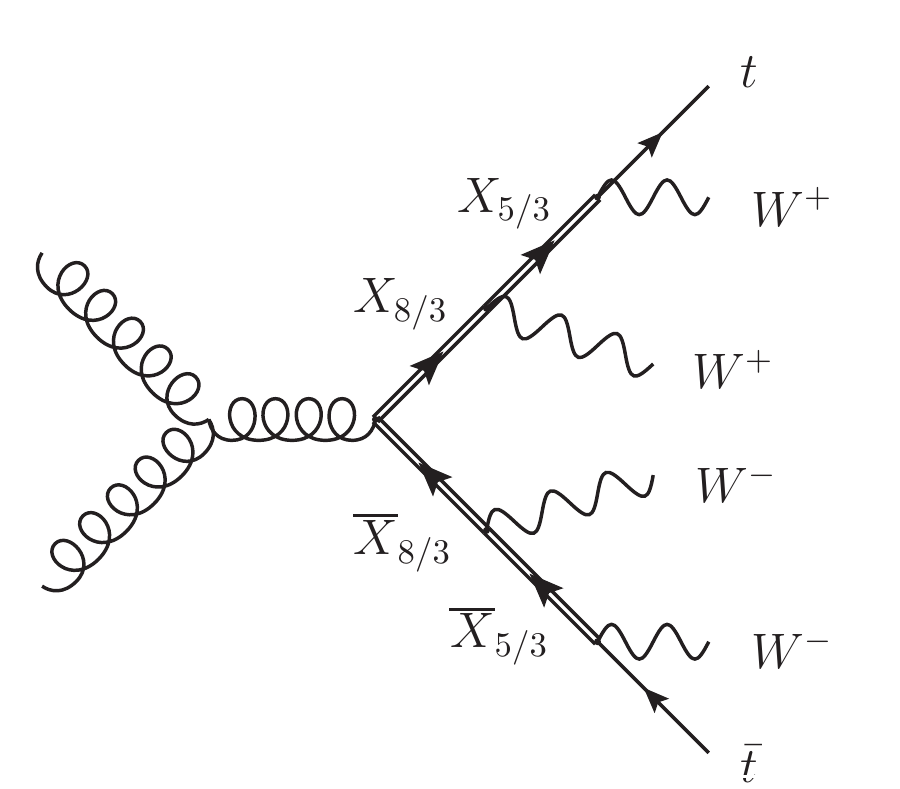}
  \caption{\emph{Pair production of the $X_{8/3}$ with a subsequent decay via contact interaction (left diagram) or via intermediate charge 5/3 state (right diagram).}}\label{fig:diagram}
\end{figure}

The $\Xet$ then decays with $\sim100\%$ probability into $W^+W^+t$, with the top quark subsequently decaying into $W^+b$, as illustrated in the left panel of Fig.~\ref{fig:diagram}. Most of the times the $W$'s decay hadronically, leading to signatures with large numbers of jets; searches for these topologies already exist in the context of microscopic black holes \cite{Chatrchyan:2013xva,atlas:bh}. As we will show in the Section~\ref{sec:fullmodel}, current sensitivity of this type of searches does not allow to put significant constraints on models with the interaction~\eq{simpleinteraction}. Although the situation is slightly better for a more complete model, in both cases multi-jet searches are not the most constraining. 

A good sensitivity to the $\Xet$ signal is achieved instead by searches for same-sign leptons. Indeed $W$'s decay leptonically about $2/9$ of the times (more if one includes leptonic $\tau$ decays) so that $\barXet\Xet$ decays produce at least two same-sign leptons ($2ssl$) approximately $1/4$ of the times -- almost three times more than charge-$5/3$ resonances. This implies that the background can be efficiently suppressed and the signal acceptance pushed to relatively large values.

 Noticeably, with a probability of $\sim 2\%$, the $\Xet$ decays into 3 leptons with the same charge ($3ssl$). This is a novel and distinctive signature, with practically vanishing background, that can potentially provide a new sensitive channel to search for $\Xet$.

In the following sections we recast one of the most up-to-date $2ssl$ searches \cite{cms53} into bounds on $M_{8/3}$, using the simplified model described above, and then we compare the sensitivity of $2ssl$  with $3ssl$ searches, both at 8 and 14 TeV, and identify the best strategy to search for charge-8/3 resonances at the LHC.

\subsection{Recast of Current  and Future Two Same-Sign Leptons Searches}
Ref.~\cite{cms53}, using 19.6\,fb$^{-1}$ of collected data, puts the strongest limit on pair produced charge $5/3$ states that decay exclusively to $W t$. 
This analysis searches for an excess of  events containing two same sign leptons ($e$ or $\mu$, including those from $\tau$ decays) and at least $N_{con}=5$ other leptons or jets. A dedicated technique is used to reconstruct top quarks and $W$-bosons from their decay products if the latter are highly boosted. The candidate leptons and jets are required to satisfy isolation criteria, minimal $p_t$ and $\eta$ cuts and the invariant mass of the leptons pairs must be away from  $M_Z$ to further suppress the $WZ$ and $ZZ$ background. On top of this, the sum of the transverse momenta of the particles in the event must be larger than $900\, \GeV$.

The search did not find any significant excess  and put a  $95\%$ C.L. lower limit of  $770\,\GeV$  on the mass of  charge $5/3$ states. This corresponds to an upper limit $N_{95} \simeq 12$ on signal events passing the selection criteria. 

The pair produced $\Xet$ can also contribute to $2ssl$ final states, and the result of Ref.~\cite{cms53} can be recast as a bound on $M_{8/3}$. We compare the allowed number of events $N_{95}$  with the expected number of accepted events from $\Xet$ decays:
\begin{equation}\label{nsignal}
N_{signal} = {\cal L}\,\text {BR}\, \epsilon_{2ssl}(M_{8/3})  \, \sigma(M_{8/3}),
\end{equation}
where ${\cal L}$ is the integrated luminosity, $\text {BR}$ is the probability (Branching Ratio) to find at least two same sign leptons among the $\barXet \Xet$ final states, $\epsilon_{2ssl}(M_{8/3})$ is the signal cut acceptance which, like the pair production cross section $\sigma(M_{8/3})$, depends on the resonance mass.

We use the values of table~\ref{hator} for $\sigma(M_{8/3})$, and compute cut acceptances performing a set of  {\sc{MadGraph~5}}~\cite{madgraph} simulations of  $\Xet$ pair production and decay (the  {\sc{MadGraph 5}} models were produced using  the {\sc{FeynRules}}~\cite{feynrules} package). We only generate parton-level events without hadronisation nor detector simulation. Using parton-level events is justified by the fact that the $\Xet$ produces a large number of jets, therefore showering would only be able to introduce small  modifications of the jets spectrum and therefore will not affect the cuts acceptances. 
A simplified version of the analysis performed in Ref.~\cite{cms53} is then applied to the generated samples, including similar jet clustering algorithms,  boosted top and $W$ tagging.

The computed acceptances $\epsilon_{2ssl}(M_{8/3})\times \text {BR}$ are presented in the second column of the left panel of table~\ref{tab:seff1} (contact interaction column). As explained in the next section, the accuracy of  our analysis can be estimated by comparing the results for the charge 5/3 states obtained using our analysis ($4^{th}$ column) and the analysis of the original paper~\cite{cms53} ($6^{th}$ column) and is  of  order 10\%. 

 From \eq{nsignal} and by requiring $N_{signal} \leq N_{95}$, we obtain, 
\begin{equation}\label{simplebound}
M_{8/3} \geq 940\,\GeV\quad \textrm{@ 95\% C.L.}\,,
\end{equation}
This bound is significantly more stringent  than the one on the charge-5/3 state ($M_{5/3}>770\,\GeV$) from the original experimental analysis~\cite{cms53}; the reasons are a factor $\sim 3$  higher branching fraction into $2ssl$ and a factor $\sim 2$  higher acceptance of the $N_{con}$ cut caused by a larger multiplicity of the $\Xet$ decay products. 
\begin{table}
\begin{center}
\begin{tabular}{ | c || c | c | c | c | c |  }
\hline
\multicolumn{6}{|c|}{8 TeV} \\
\hline
$\begin{array}{c} M \\ \textrm{[GeV]} \ea$ & $\begin{array}{c} Q=\tfrac{8}{3} \\ \textrm{(contact)} \ea$& $\begin{array}{c} Q=\tfrac{8}{3} \\ (\textrm{via }\tfrac{5}{3}) \ea$  &  $\begin{array}{c} Q=\tfrac{5}{3} \\   \textrm{ } \ea$ &  $\begin{array}{c} Q=-\tfrac{1}{3} \\ \textrm{ } \ea$ &  $\begin{array}{c} Q=\tfrac{5}{3} \\ \textrm{(from \cite{cms53})} \ea$ \\  
  \hline
    \hline
 600   & $ 51$ &  101   & $ 11$  &  15 &  13  \\
 800    & $ 97$&  108   & $ 19$  &  22 &  23 \\
 1000  & $124$&  114   & $ 23$  &  24 &  26 \\
 1200 & $133$ &  119   & $ 24$  &  25 & -- \\
 1400 & $138$ &  122   & $ 24$  &  25 & -- \\
 1600  & $139$&  125   & $ 24$  &  24 & -- \\
  \hline
\end{tabular}
\hspace{0.2cm}\begin{tabular}{ | c || c | c | c | c |  }
\hline
\multicolumn{5}{|c|}{14 TeV} \\
\hline
$\begin{array}{c} M \\ \textrm{[GeV]} \ea$  & $\begin{array}{c} Q=\tfrac{8}{3} \\ \textrm{(contact)} \ea$& $\begin{array}{c} Q=\tfrac{8}{3} \\ (\textrm{via }\tfrac{5}{3}) \ea$  &  $\begin{array}{c} Q=\tfrac{5}{3} \\   \textrm{ } \ea$ &    $\begin{array}{c} Q=\tfrac{5}{3}  \\ \textrm{(from \cite{cms14})} \ea$ \\  
  \hline
    \hline
 1000  & 22.7  &  76.6   &  5.53   &    7.10  \\
 1200  & 51.9  &  91.9   &  13.7   &   12.3  \\
 1400  & 83.1  &  103    &  17.6   &   15.0  \\
 1600  & 114   &  115    &  21.5   &   16.9  \\
 1800  & 128   &  118    &  23.7   &   16.8  \\
 2000  & 136   &   119   &  23.6   &   16.1  \\
  \hline
\end{tabular}\end{center}
\caption{Acceptance $\epsilon_{2ssl}(M_{8/3}) $ for the cuts of the \emph {2ssl} analysis of Ref.~\cite{cms53} at 8 TeV (left panel) and of  Ref.~\cite{cms14} at 14 TeV (right panel), multiplied by BR$\times 10^3$, for top partners of different electric charges $Q$; numbers include the BR's of the $W$ bosons but assume that all the $5/3$ states decay exclusively to $t+W$. The acceptance for the $X_{8/3}$ is given separately for two possible decay channels: with intermediate $X_{5/3}$ or $Y_{5/3}$, \eq{5383}, and via contact interactions with a $d$-symbol, \eq{dd83}. The last columns corresponds to  the original analyses~\cite{cms53,cms14}.
Given that their decays have similar topology, at 14 TeV, efficiencies for the charge -1/3 states are taken equal to the ones of the 5/3.
}
\label{tab:seff1}
\end{table}

Similarly, the reach of the 14 TeV LHC on the $\Xet$ mass can be estimated by recasting the exploratory analysis of Ref.~\cite{cms14}, tailored for charge-5/3 states. The main difference with respect to the 8 TeV analysis of Ref.~\cite{cms53}, is a harder cut on the transverse momenta.  The  efficiencies for  $\Xet$ and the 14 TeV cuts, are reported in the second column of the right panel of  table~\ref{tab:seff1} and the mass reach is illustrated in  Fig.~\ref{fig:exclusion_1}.
Again, we can judge the accuracy of our study by comparing our  efficinecy for charge-5/3 states, with those of Ref.~\cite{cms14} ($4^{th}$ and $5^{th}$ columns of the right panel). The two analyses differ here by at most 20\% at low masses and by up to  47\% at 2 TeV. This means that our analysis, while still providing a good estimate of the experimental sensitivity,  misses some effects, likely related to the high boost and the collinearity  of the decay products. Nevertheless, in the case of $\Xet$, the energy is distributed among a larger number of particles which are consequently less boosted than for the charge 5/3, implying that the distortion between a realistic analysis and ours will be smaller. Another factor that reduces the sensitivity to high boosts, is the collinearity between the $b$ and the eventual lepton in the top-quark decay, which compromises the ability to single out the lepton. This effect, affects in a bigger proportion searches for charge-5/3 states, which produce at most two leptons (and if one is lost do not pass the $2ssl$ cut), than $\Xet$ searches, which are most likely to produce non-collinear leptons.

\subsection{Dedicated Search for $X_{8/3}$ in Three Same-Sign Leptons}\label{sec:3ssl}

In this paragraph, we analyze the possibility to construct a different, dedicated, experimental search to test the production of  charge $8/3$ states: with $3ssl$ final states. This analysis would certainly be necessary if a $2ssl$ signal is ever observed, in order to distinguish between the $\Xet$ and other resonances with $2ssl$ decays, but it can also potentially be used to search directly for the $\Xet$. In what follows, we compare the sensitivity of a $3ssl$ search w.r.t. the $2ssl$ one, in order to establish their relative exclusion potential.

The great advantage of the $3ssl$ channel is that the background is practically vanishing. The $3ssl$ events in the SM can originate as genuine $3ssl$ signals or as  $2ssl$ events in which the charge of one of the extra leptons has been misidentified, or a  jet has been taken for a lepton. 
The former can be predicted from theory, the dominant contributions coming from $ZZZ$, $WZZ$ and $WWZ$ events, and their rate is about a factor $\sim \alpha_{\rm em}$ smaller than for the $WZ$ and $ZZ$ backgrounds affecting $2ssl$ searches (see Refs.~\cite{cms53,cms53OLD}). The $ZZZ$ and $WZZ$ events, together with 
contributions from $\bar t t Z$, are efficiently eliminated with a $Z$ veto, requiring the invariant mass of any two leptons to be off the Z-pole. On the other hand, the part from $WWZ$, and $\bar t t W$, is less sensitive to the Z veto, but is penalized by requiring a large number $N_{con}$ of extra hard constituents in the event, since these events are not typically accompanied by several hard jets.

Leptons with misidentified charge, on the other hand, correspond to a genuine $2ssl$ background (dominantly $WZ$ and $ZZ$)  with extra misidentified leptons.  While the probability to misidentify muons is negligible, the electrons/positron misidentification probability is estimated as $P_{misid}=5.89\times 10^{-4}$ \cite{cms53OLD}. The $Z$ veto is also efficient in this case.

 Finally, backgrounds due to jet misidentification are typically extracted using data-driven techniques which lie beyond the reach of our analysis. Nevertheless, this source of background is efficiently eliminated by requiring a large number of final states~\cite{cms53OLD}. 

In order to suppress these background most efficiently, while preserving the signal, we apply the following selection cuts: 
\begin{itemize}
\renewcommand{\labelitemi}{$\blacktriangleright$}

\item Reconstruction criteria:

\renewcommand{\labelitemi}{$\circ$}

\item Leptons ($e$ and $\mu$) are required to have $p_T(l)>30\,\GeV$ and pseudorapidity $|\eta(l)|<2.4$. They should also satisfy the following isolation criterium: sum of the $p_T$ of the objects inside a cone with a radius $\Delta R = 0.3$ around a lepton candidate should not exceed 15\% (20\%) of the electron (muon) $p_T$.

\item Top jets are reconstructed using the Cambridge-Aachen clustering algorithm~\cite{CA} with a distance parameter $R=0.8$, and are required to have a $p_T>200\,\GeV$, $|\eta|<2.4$, invariant mass  $m_{inv}\in[140,250]\,\GeV$, at least 3 constituent subjets and a minimal pair-wise mass of the constituents of at least $50\,\GeV$.

\item W jet candidates are also reconstructed using the Cambridge-Aachen clustering algorithm with $R=0.8$ and with requirements $p_T>200\,\GeV$, $|\eta|<2.4$, $m_{inv}\in[60,130]\,\GeV$ and must consist of two subjets.

\item Jets which are not identified as boosted tops or $W$'s are clustered using anti-$k_T$ algorithm~\cite{antikt} with $R=0.5$ and are required to have $p_T>30\,\GeV$ and $|\eta|<2.4$.

\item Any jet must be separated from the reconstructed leptons by at least $\Delta R = 0.3$ and from other jets by $\Delta R = 0.8$.

\renewcommand{\labelitemi}{$\blacktriangleright$}

\item Event selection:

\renewcommand{\labelitemi}{$\circ$}
\item 3 same sign leptons ($e$ or $\mu$).
\item Z and quarkonia veto: $M(l l)>20\,\GeV$ for any pair of leptons, $M(\mu^+ \mu^-)\notin [76,106]\,\GeV$ for opposite-sign muons and $M(e e)\notin [76,106]\,\GeV$ for any pair of electrons.   
\item A minimal number of constituents $N_{con}=3$ apart from $3ssl$ (this includes other leptons and jets candidates, with top jets counted as three and $W$'s  as two constituents).
\end{itemize}

We simulated the most relevant backgrounds using {\sc{MadGraph 5}}  and compared the efficiency of the cuts  described above.  For 100 fb$^{-1}$, at 8 TeV (14 TeV), the number of $3ssl$ background events from $WZ$ and $ZZ$ with a misidentified lepton, is approximately 2 (5); this reduces below sensitivity after the Z veto. This is true also for genuine $3ssl$ contributions from $WZZ$ and $ZZZ$, which are reduced from about 3 (4) events to $\sim 0.2 (0.3)$ and are rendered negligible by a further $N_{con}$ cut. The $3ssl$ contribution from $\bar t t W$ (and also the one from $WWZ$) is very small (of order 0.2 (0.1)) and can be neglected.

On the contrary, the signal is almost unaffected by these selection cuts, as shown in the upper panels of table~\ref{tab:seff3ssl}, where we summarize  the cut acceptances (including branching ratios) $\epsilon_{3ssl}(M_{8/3})\times $BR for different masses at 8 and 14 TeV, obtained from the same simulation as in the previous paragraph. 

\begin{figure}[!t]
    \begin{center}
    \hspace{-1.cm}
      \includegraphics[width=0.47\textwidth]{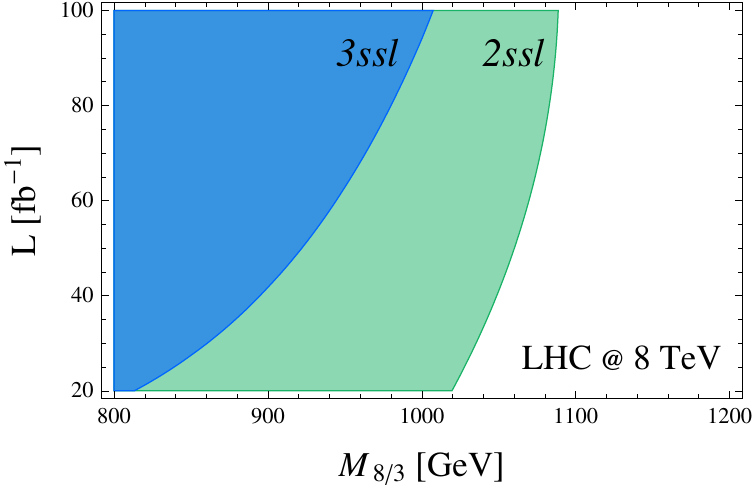}
      \includegraphics[width=0.47\textwidth]{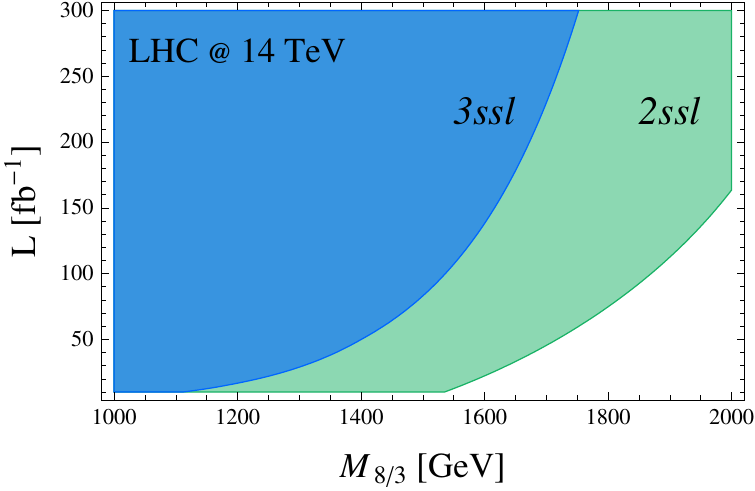}
  \end{center}
\caption{\emph{A comparison of expected excluded masses of the charge $8/3$ state for the $2ssl$ (green) and $3ssl$ (blue) search channels in the simplified model for different integrated luminosities for 8 TeV (left panel) and 14 TeV (right panel).}}\label{fig:exclusion_1}
\end{figure}

In order to estimate the excluding power of the $3ssl$ we performed a statistical analysis assuming that the observed signal equals to background, i.e. there is no excess, given that at present no experimental data is available for $3ssl$ channel. Under this assumption, the hypothesis predicting more than $N_{95}^{3ssl}=3$ events is excluded with a 95\% CL. Then, using \eq{nsignal}, we estimate the bound on the $\Xet$ mass depending on the integrated luminosity, that we report in Fig.~\ref{fig:exclusion_1}. Notice that a fair comparison between $2ssl$ and $3ssl$ requires that the number of registered events equals to the estimated background in both cases. In this hypothetical situation, the $2ssl$ search leads to a bound on the number of $2ssl$ events stronger than the actual one: at 8 TeV $N_{95}^{2ssl}\simeq 7$ (in the real data a small excess over the expected number of events was observed). Then, the $2ssl$ analysis would have been able to constrain $\Xet$ up to  $M_{8/3} \geq 1010\,\GeV$ (at 8 TeV and $\sim 20$ fb$^{-1}$). 

As we can see from  Fig.~\ref{fig:exclusion_1}, the 3ssl channel would not be able to overpass $2ssl$ neither for the 8 TeV LHC with increased integrated luminosity, nor for 14 TeV experiments. 
The smallness of the background can not compensate a great drawback of the $3ssl$ search: the small BR$\sim 2\%$ into three same-sign final state leptons reduces the signal acceptance by roughly a factor of 10.
We conclude that, although the $3ssl$ search remains an important discriminant for these models in case of discovery, its sensitivity is not competitive with $2ssl$ searches.

\begin{table}[top]
\begin{center}
\begin{tabular}{c}
\begin{turn}{90}  Contact Interaction \end{turn}\\
\end{tabular}
\begin{tabular}{ | c || c | c | c |    }
\hline
\multicolumn{4}{|c|}{8 TeV}\\
\hline
 Mass, GeV    & 3ssl & $M_{ll}$ & $N_{con}\geq3$    \\  
 \hline
\hline
600             &  10.4 & 7.88 & 7.52 \\
800             &  9.86 & 8.08 & 7.90 \\
1000           & 11.4 & 9.78 & 9.61 \\
1200           & 12.1 & 10.4 & 10.3 \\
1400           & 12.2 & 11.0 & 10.9 \\
1600           & 12.2 & 11.1 & 10.9   \\
\hline
\end{tabular}
\hspace{1cm }
\begin{tabular}{c}
\begin{turn}{90}  Contact Interaction \end{turn}\\
\end{tabular}
\begin{tabular}{|c || c | c | c |  }
\hline
\multicolumn{4}{|c|}{14 TeV}\\
\hline
Mass, GeV    & 3ssl & $M_{ll}$ & $N_{con}\geq{3}$   \\  
 \hline
\hline
1000 & 11.6 & 9.98 & 9.71 \\
1200 & 12.5 & 11.0 & 10.9 \\
1400 & 13.0 & 11.8 & 11.7 \\
1600 & 13.6 & 12.6 & 12.5 \\
1800 & 13.6 & 12.7 & 12.6 \\
2000 & 12.5 & 11.9 & 11.7  \\
\hline
\end{tabular}\end{center}
\begin{center}
\begin{tabular}{c}
\begin{turn}{90}  Via charge-5/3 states \end{turn}\\
\end{tabular}
\begin{tabular}{ | c || c | c | c |    }
\hline
\multicolumn{4}{|c|}{8 TeV}\\
\hline
 Mass, GeV    & 3ssl & $M_{ll}$ & $N_{con}\geq3$    \\  
 \hline
\hline
600             &  8.11 & 7.14 & 6.72 \\
800             &  9.86 & 7.82 & 7.81 \\
1000           & 8.92 & 7.74 & 7.53  \\
1200           & 10.4 & 9.03 & 8.85  \\
1400           & 9.37 & 7.57 & 7.56  \\
1600           & 10.4 & 9.69 & 9.64   \\
\hline
\end{tabular}
\hspace{1cm}
\begin{tabular}{c}
\begin{turn}{90}  Via charge-5/3 states \end{turn}\\
\end{tabular}
\begin{tabular}{|c || c | c | c |  }
\hline
\multicolumn{4}{|c|}{14 TeV}\\
\hline
Mass, GeV    & 3ssl & $M_{ll}$ & $N_{con}\geq3$   \\  
 \hline
\hline
1000 & 10.3 & 8.69 & 8.56 \\
1200 & 9.71 & 8.47 & 8.39 \\
1400 & 10.7 & 9.72 & 9.62 \\
1600 & 11.8 & 10.9 & 10.8 \\
1800 & 11.0 & 10.1 & 10.1 \\
2000 & 10.5 & 9.74 & 9.56 \\
\hline
\end{tabular}\end{center}
\caption{\emph{Acceptance of the cuts times BR $\times 10^{3}$ for 3ssl from the $X_{8/3}$ in the simplified model (upper panels) and for decays via an off-shell charge-5/3 state (lower panels).}}
\label{tab:seff3ssl}
\end{table}

\section{Charge $8/3$ Resonances in Composite Higgs Models}\label{sec:fullmodel}
As mentioned above, the best motivated scenarios for states with charge $8/3$ are composite Higgs models, where $\Xet$ arises as a composite resonance, as we now briefly summarize.  
In the simplest composite Higgs models~\cite{Agashe:2004rs,Contino:2006qr}, a new strongly interacting sector is postulated, which possesses a global $SO(5)\times U(1)_X$ symmetry in the UV, broken dynamically  to $SO(4)\times U(1)_X$ in the IR, such that the Higgs arises as the pseudo-Nambu-Goldstone boson parametrizing the coset $SO(5)/SO(4)$. Then we can describe it with the Goldstone boson matrix $U(\Pi)=\exp\left(i\,\sqrt{2}\, \Pi^{\hat a} \,T^{\hat a}/f\right)$, where $T^{\hat a}$ are the broken generators, with $\hat a = 1...4$, $\Pi$fields correspond to the Higgs doublet 
and the scale $f$ is the analog of the pion decay constant.  Using the notation of Refs.~\cite{Contino:2006qr,Agashe:2004rs}, in the unitary gauge we have
\begin{align}
U=\,& \(\bry{ccc|cc}
& & \vspace{-3mm}& & \\
 & \mathbb{I}_{3} & & &  \\
  & & \vspace{-3mm}& &  \\ \hline
   & & & \cos \frac{\langle h \rangle + h}{f} & \sin \frac{\langle h \rangle + h}{f} \\
    & & & -\sin \frac{\langle h \rangle + h}{f} & \cos \frac{\langle h \rangle + h}{f} \ery\)\,,
\end{align}\\
where $h$ is the Higgs boson and the Higgs VEV $\langle h \rangle$ is fixed to reproduce the correct mass of the SM gauge bosons by $f \sin {\langle h \rangle \over f} = v \equiv 246$~GeV. The parameter $\xi$, defined as
\be
\xi \equiv \left({v \over f} \right)^2<1 \, ,
\ee 
characterises the separation between the electroweak scale and the scale of the strong sector resonances and is expected to be at most as large as $\sim 0.2$ to allow the model to pass the constraints imposed by Electroweak Precision Tests~\cite{ewpt}. 

A subgroup $SU(2)_L\times U(1)_Y$ of  the residual global symmetry $\mathcal{H}=SO(4)\times U(1)_X\simeq SU(2)_L\times SU(2)_R\times U(1)_X$, is weakly gauged and corresponds to the SM gauge group, with hypercharge corresponding to the diagonal combination 
\begin{equation}\label{EMrelation}
Y=X+T_R^3.
\end{equation}
Under these assumptions, the Higgs doublet  is fixed to transform as a $({\bf 2},{\bf 2})_0$ under $\mathcal{H}$, where the subscript denotes its $X$-charge. For the $SO(5)$ and $SO(4)$ generators we adopt the  conventions of Ref.~\cite{DeSimone:2012fs}.

From the point of view of flavor physics,  couplings between SM fermions and the strong sector prefer the partial compositeness \cite{Kaplan:1991dc} paradigm~\footnote{For the discussion of compatibility of partial compositeness with flavour constraints we refer the reader to Refs.~\cite{KerenZur:2012fr,Barbieri:2012tu}.}. Under this assumption, each SM fermion couples linearly to some fermionic operator $\mathcal{O}$ of the strong sector. As an example for left-handed SM doublets $q_L$ in the UV we can write
\begin{equation}
\mathcal{L}=y \, \bar q_L^\alpha \, \Delta_{\alpha,I_{\mathcal{O}}} \, \mathcal{O}^{I_{\mathcal{O}}}+{\rm h.c.}\equiv y\,(\bar Q_L)_{I_{\mathcal{O}}}\, \mathcal{O}^{I_{\mathcal{O}}}+{\rm h.c.},
\label{partcompmix}
\end{equation}
where $I_{\mathcal{O}}$ denotes the indices of the operator $\mathcal{O}$ transforming in a representation $r_{\mathcal{O}}$ of  $SO(5)\times U(1)_X$;
 with $(Q_L)_{I_{\mathcal{O}}} =  \Delta^{*}_{I_{\mathcal{O}}, \alpha} \,q_L^\alpha$ we denote the embedding of $q_L$ into a full representation of $SO(5)$~\cite{Contino:2010rs}. The mixing of \eq{partcompmix}) breaks explicitly the $SO(5)$ symmetry ($y \, \Delta$ is the spurion of such a breaking), which protects the Higgs mass, therefore a sufficiently light Higgs would in general require a small value of the breaking strength $y$. 
 The representation $r_{\mathcal{O}}$, that the fermions couple to, is not fixed by the low-energy dynamics, but depends on details of the UV realization. Minimal models where the third family quarks couple to operators with  $r_{\mathcal{O}}={\bf 5_{2/3}}$ (here {\bf 5} denotes the $SO(5)$ representation, while the subscript denotes its $U(1)_X$ charge) have been extensively studied~\cite{Contino:2008hi,DeSimone:2012fs}.
 Here we are interested in models where the right-handed top arises as a composite resonance from the strong sector, while the third family quark doublet  couples to operators in the symmetric traceless representation $r_{\mathcal{O}}={\bf 14_{2/3}}$ \cite{Pomarol:2012qf,Pappadopulo:2013vca,Panico:2012uw,DeSimone:2012fs} through the embedding
\be\label{spurions}
\bry{l}
\mathcal{Q}_{L}= \,\dst \f{1}{\sqrt 2} \begin{pmatrix} & & & & i b_{L} \\
& & & & b_{L} \\
& & & & i t_{L} \\
& & & & - t_{L} \\
i b_{L}& b_{L} & i t_{L} & - t_{L} &   \\ \end{pmatrix}\, ,
\ery
\ee
which makes the charge -1/3 sector of the theory invariant under the left-right parity protecting the $Zb_L b_L$ vertex from large tree-level corrections~\cite{Agashe:2006at}.  
Then, at low energies, these operators can excite the states transforming as ${\bf 1_{2/3}}$, ${\bf 4_{2/3}}$ and ${\bf 9_{2/3}}$ under the $SO(4)\times U(1)_X$, which make up a full ${\bf 14_{2/3}}$ of $SO(5)\times U(1)_X$. The phenomenology of the first two was already studied in Ref.~\cite{DeSimone:2012fs}; in this work we concentrate, instead, on the phenomenology of a multiplet of resonances $\Psi$ in the ${\bf 9_{2/3}}$, equivalently $({\bf 3,3})_{2/3}$ of $SU(2)_L\times SU(2)_R\times U(1)_X$ which, in holographic models, has been shown to be the lightest multiplet~\cite{Pappadopulo:2013vca}. Its components can be further divided according to how they transform under the  SM gauge group: three triplets of $SU(2)_L$ with charges given by \eq{EMrelation} \cite{Pappadopulo:2013vca,Montull:2013mla},
\begin{equation}\label{decomposition}
\Psi\supset \{ {\Xet,\Xft,\Xtt} \},\, \{ {\Yft,\Ytt,\Yot} \} ,\, \{ {\Ztt,\Zot,\Zft} \},
\end{equation}
separated according to their $T^3_R=+1,0,-1$ eigenvalues, with subscripts corresponding to electric charges. The full SO(4) nine-plet $\Psi$ can be written as (the representation is symmetric and elements in the upper diagonal have been omitted for clarity)

\be
{1 \over 2}
\left(
\begin{array}{ccccc}
 -\Xet+\Ytt-\Zft &   &   &   \\
  i \Zft- i \Xet &  {\Xet}+{\Ytt}+{\Zft} &   &    \\
 \frac{\Xft}{\sqrt{2}}-\frac{\Yot}{\sqrt{2}}+\frac{\Yft}{\sqrt{2}}-\frac{\Zot}{\sqrt{2}} & \frac{i \Xft}{\sqrt{2}}+\frac{i \Yot}{\sqrt{2}}+\frac{i \Yft}{\sqrt{2}}+\frac{i \Zot}{\sqrt{2}} & -{\Xtt}-{\Ytt}-{\Ztt}   &   \\
 -\frac{i \Xft}{\sqrt{2}}+\frac{i \Yot}{\sqrt{2}}+\frac{i \Yft}{\sqrt{2}}-\frac{i    \Zot}{\sqrt{2}} & \frac{\Xft}{\sqrt{2}}+\frac{\Yot}{\sqrt{2}}-\frac{\Yft}{\sqrt{2}}-\frac{\Zot}{\sqrt{2}} &  i \Xtt- i \Ztt & {\Xtt}-{\Ytt}+{\Ztt}  
\end{array}
\right)
\ee

Our goal, in what follows, will be to study the experimental constraints on these models, as can be extracted mainly from the direct searches for composite fermions. Focusing on the state with charge $8/3$, we will also highlight the extent to which the simplified model discussed in section~\ref{sec:simplemodel} can be considered as a good approximation of this more realistic setup.

\subsection{Spectrum and Decays} 

The CCWZ construction \cite{Coleman:1969sm} allows us to build the most general effective Lagrangian describing the leading interactions between a pNGB Higgs,  the strong sector resonances $\Psi$  in an $SO(4)\times U(1)_X$ multiplet ${\bf 9_{2/3}}$ and the third-family SM quark doublet embedded in a ${\bf 14_{2/3}}$ of $SO(5)\times U(1)_X$:
\begin{align}\label{LagCCWZ}
\mathcal{L}=&\,i\,\bar q_L\,\Dsl\, q_L+i\,\bar t_R \,\Dsl \, t_R\nonumber\\
&+ c_{1} y f \, \bar Q_L^{I,J}\,U_{I,i}\,U_{J,j}\,\Psi_R^{i,j}+y f \, \bar Q_L^{I,J}\,U_{I,5}\,U_{J,5} \, t_R + \textrm{h.c.} \\
 &+ \frac{f^2}{4}d_\mu^id^{\mu,i}+\left[i \, \bar\Psi^{i,j}(\Dsl+2i\,\slashed{e}^aT_a^{j,k})\Psi^{k,i}-M \,\bar \Psi \Psi\right] 
+\frac{c_{2}}{M_*}\bar\Psi_L^{i,j}\,d_\mu^i \,d^{\mu, j}\,t_R
\,,\nonumber
\end{align}
where  $i$ and $I$ are $SO(4)$ and $SO(5)$ indices respectively, and $a=1,...,6$ enumerates the $SO(4)$ generators~$T^a$. Here the second line includes the linear mixings between the elementary fermions and the strong sector resonances, as implied by partial compositeness. The third line describes the strong sector alone in terms of the $d$ and $e$ symbols, defined at leading order in the $h/f$ expansion~\footnote{The full expressions can be found for example in the Ref.~\cite{DeSimone:2012fs}} as
\begin{equation}
d_\mu^i=\frac{\sqrt{2}}{f}(D_\mu h)^i+{\cal O} (h^3) \quad\textrm{and}\quad e_\mu^a=-g A_\mu^a-\frac{i}{f^2}(h{\lra{D}_\mu}h)^a+{\cal O} (h^4);
\end{equation}
 this guarantees that this theory originates from the spontaneous symmetry breaking of $SO(5)\to SO(4)$.
  The  mass of $\Psi$ is denoted $M$, while $M_*$ is the mass of  heavier composite resonances that have been integrated away.
 The dimensionless coefficients $c_1$ and $c_2$ are expected of order unity from  power counting arguments~\cite{Giudice:2007fh}. The model described above does not possess  sufficient structure to be able to predict some of the key quantities related to EWSB, such as the Higgs VEV or mass, this additional structure can be added for example following the prescription of the Ref.~\cite{Panico:2011pw} or \cite{Pomarol:2012qf,marzoccaserone}.  
   \begin{figure}[!t]
    \begin{center}
      \includegraphics[width=0.72\textwidth]{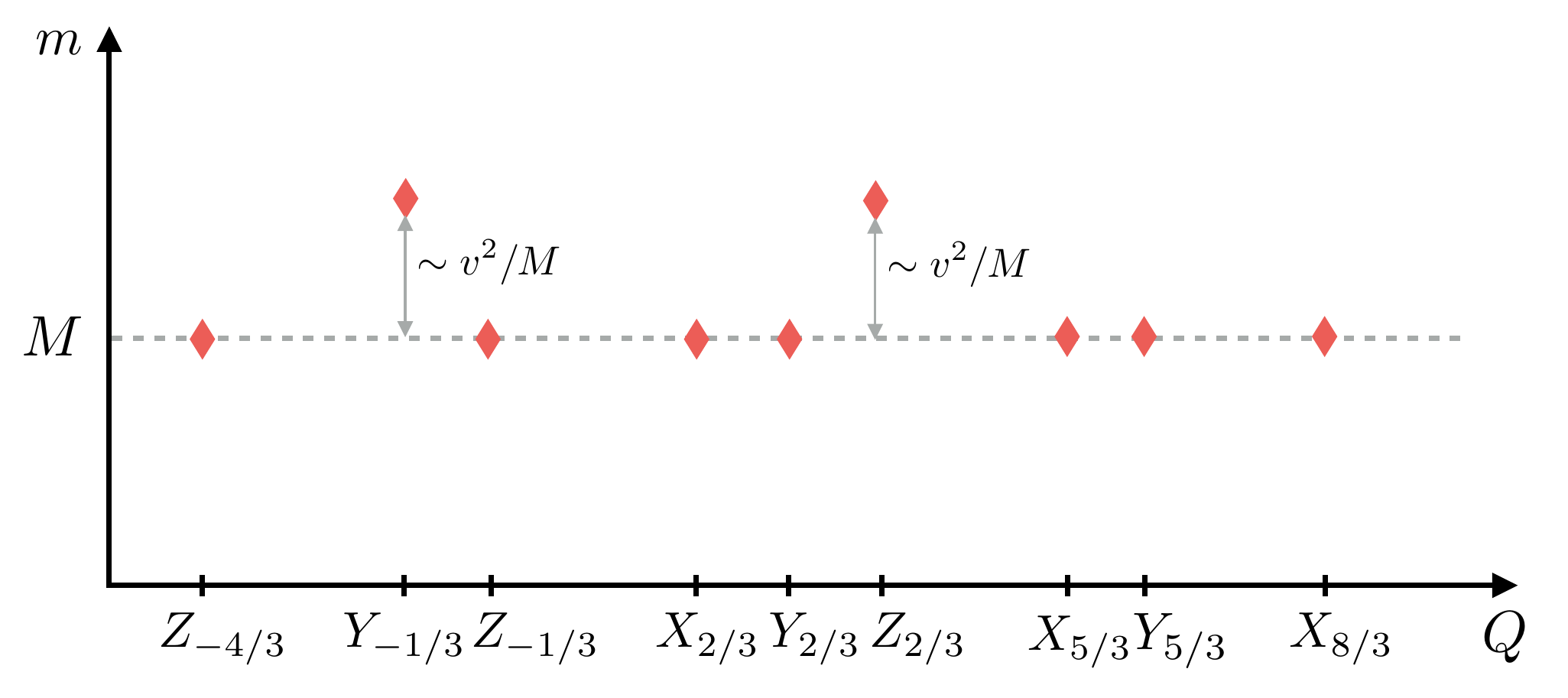}
  \end{center}
\caption{\emph{Mass spectrum of the new heavy states present in a model with $({\bf 3},{\bf 3})_{2/3}$.}}\label{fig:masses}
\end{figure}
 
States with electric charge $8/3,5/3$ and $-4/3$ do not mix with the SM, at tree level their mass matrices are diagonal and they are all degenerate with mass $M$. States with charge $2/3$ or $-1/3$, on the other hand, mix with the elementary top and bottom quarks, lifting the degeneracy. In the mass eigenstate basis one finds, for the charge $2/3$ sector, a light state corresponding to the top quark (its mass $\sqrt{2} m_t\simeq y f\sin(2v/f)$ fixes the coupling $y$), two degenerate states (called $\Xtt$, $\Ytt$ in the following) with mass $M$, and one slightly heavier state ($\Ztt$) with mass $\sim M+{5 \over 4} c_1^2 y^2v^2/M$. 
In the charge $-1/3$ sector we have a massless bottom quark (for simplicity we have not included $b_R$ in \eq{LagCCWZ}, which does not affect the phenomenology of the top partners), and heavy $\Yot $ and $\Zot$ partners with masses  $\sim M+c_1^2 y^2 v^2 / M$ and $M$ respectively.
 Such tiny splitting $\sim v^2$ (compared to $f^2$ in the case of the ${\bf 4_{2/3}}$ ~\cite{DeSimone:2012fs}) is a peculiarity of the ${\bf 9_{2/3}}$ and is due to the fact that in order to construct an $SU(2)_L$ singlet mass mixing term out of the triplets of \eq{decomposition} and the SM doublet $q_L$, an insertion of the Higgs doublet is required.

In summary, we expect an almost degenerate mass spectrum of resonances, as illustrated in Fig.~\ref{fig:masses}. 
In this situation the phenomenology of top partners is rather simple: cascade decays are kinematically disfavored and two-body decays into a SM fermion and a gauge boson dominate, when allowed.\footnote{For completeness we have extended our analysis into the $M_{5/3}< M_{8/3}$ region and found that (for $M_{5/3}>770$ GeV, as implied by direct searches~\cite{cms53}) the efficiency is smaller by at most $\approx 7\%$ w.r.t. the contact interaction case; this result implies that the bounds that we find in this article hold also when cascade decays are allowed.}
Up to corrections of order $g/y$, the decays of the states belonging to $Y$ and $Z$ triplets are controlled by the term proportional to $c_1$ in \eq{LagCCWZ}.
We estimate the size of these couplings using the equivalence theorem and consider interactions with the goldstone bosons $\phi^{0,\pm}$ rather than with gauge bosons $Z,W^\pm$ (this is equivalent to consider only the interactions of the longitudinal modes of the gauge bosons and neglect the transverse ones, an approximation motivated by the $g/y$ expansion). As a result we can treat all three-particle interactions, including the ones with the Higgs, on the same footing and compute the branching ratios directly, based on the coupling values. At the leading order in $\xi$, the relevant interactions of the mass eigenstates are~\footnote{The states $\Xft$ and $\Yft$ are degenerate at tree level, but split by loop effects. The leading such effects, coming from the Yukawa with the elementary top quark, align the states such that \eq{eq:lagrgold1} holds. Nevertheless, interactions with transverse elementary gauge fields, can introduce corrections of  order ${\cal O}(g^2 \xi)$. Similar arguments apply  to  other degenerate states: true mass eigenstates will differ from the ones used in the  eq.~(\ref{eq:lagrgold1}) by at most a rotation proportional to $\xi$; this will not affect the discussion which follows.}
\bea
{\mathcal L}&\supset& - c_1 y \, \bar{t}_L \left[ \sqrt{2}\, \phi^- Y_{5/3} +\phi^+ Y_{1/3}+\phi^+ Z_{1/3}+{1\over \sqrt{10}}\left(4 i\, \phi^0 Y_{2/3}+(3 i\, \phi^0+5\, h)Z_{2/3}\right)\right]\nn\\
&&+c_1 y\, \bar{b}_L \left[  {2 \over \sqrt{5} }\, \phi^- Y_{2/3} -{1\over \sqrt 5}\, \phi^- Z_{2/3}-2\, \phi^+ Z_{-4/3}+\sqrt{2}\, h Y_{-1/3} +   i \sqrt{2}\, \phi^0 Z_{-1/3}\right]\nn\\
 &&- c_1 \xi\frac{y}{\sqrt{2} }\, \left[(h+i\phi^0)\bar{t}_L X_{2/3} -   {\sqrt{2} }\, \phi^-\bar{b}_L X_{2/3} \right]+\textrm{h.c.}\, ,
\label{eq:lagrgold1}
\eea
The extra $\xi$ suppression for members of the $X$ group is due to the fact that they mostly consist of states with the right isospin $T_R^3(X)=+1$ and need therefore at least three insertions of the Higgs (which is  a doublet of $SU(2)_R$) to couple with the SM fermions, whose right isospin is $T_R^3(q)=-1/2$.
Couplings of $\Xft$ to the top quark and $\phi^+$ are present at subleading orders in $\xi g^2 / y^2$, in addition $\Xft$ couples to the transverse components of the $W$, therefore $\Xft$ is expected to decay with probability $\sim1$ into $W t$.

For charge conservation,  there are no two-body decays of $\Xet$ into SM fields; its dominant interactions come from the covariant derivative
\begin{equation}\label{5383}
\mathcal{L}\supset g\,\barXet \slashed W^+  \Xft+g \xi\,\frac{3}{4}\,\barXet \slashed W^+ \Yft+\textrm{h.c.}
\end{equation}
and from the effective interaction in the last term of  \eq{LagCCWZ},
\begin{equation}\label{dd83}
\mathcal{L}\supset -\xi \, \frac{c_{2}\, g^2}{2 M_*}\,\barXet \, W^+ W^+ \, t_R +\textrm{h.c.}
\end{equation}
We can now easily estimate the branching ratios, ignoring the masses of the final states:
\begin{center}
\begin{tabular}{lll}
&&\\
BR$(\Yft \to  t_L W^+)\simeq 100\%$ & BR$(\Yot \to t_L W^-)\simeq 33\% 		$	& BR$(\Ytt \to t_L Z)\simeq 66\%   $ \nn\\
						&BR$(\Yot \to b_L h)\simeq 66\%$  &	 BR$(\Ytt \to b_L)\simeq 33\%$         \nn\\
						&&\\
BR$(\Zft \to b_L W^-)\simeq 100\%	$&BR$(\Zot \to t_L W^-)\simeq 33\%$\ &	 BR$(\Ztt \to t_L h)\simeq 70\%$                   \nn\\
						&BR$(\Zot \to b_L Z)\simeq 66\%$	&	BR$(\Ztt \to t_L Z )\simeq 25\%$   \nn\\
												&		&  BR$(\Ztt \to b_L W^+)\simeq 5\%$\nn\\
														&&\\
BR$(\Xet\to t_R W^+W^+)\simeq 100\%$\hspace{0.5cm}		&			BR$(\Xft\to t_L W^+)\simeq 100\%$&\\
&&\\				
\end{tabular}
\end{center}
Here we didn't list the BR's of the $\Xtt$ because they can not be reliably computed without accounting for loop effects in the mass matrix. 

\begin{figure}[!t]
    \begin{center}
    \hspace{-1.cm}
      \includegraphics[width=0.47\textwidth]{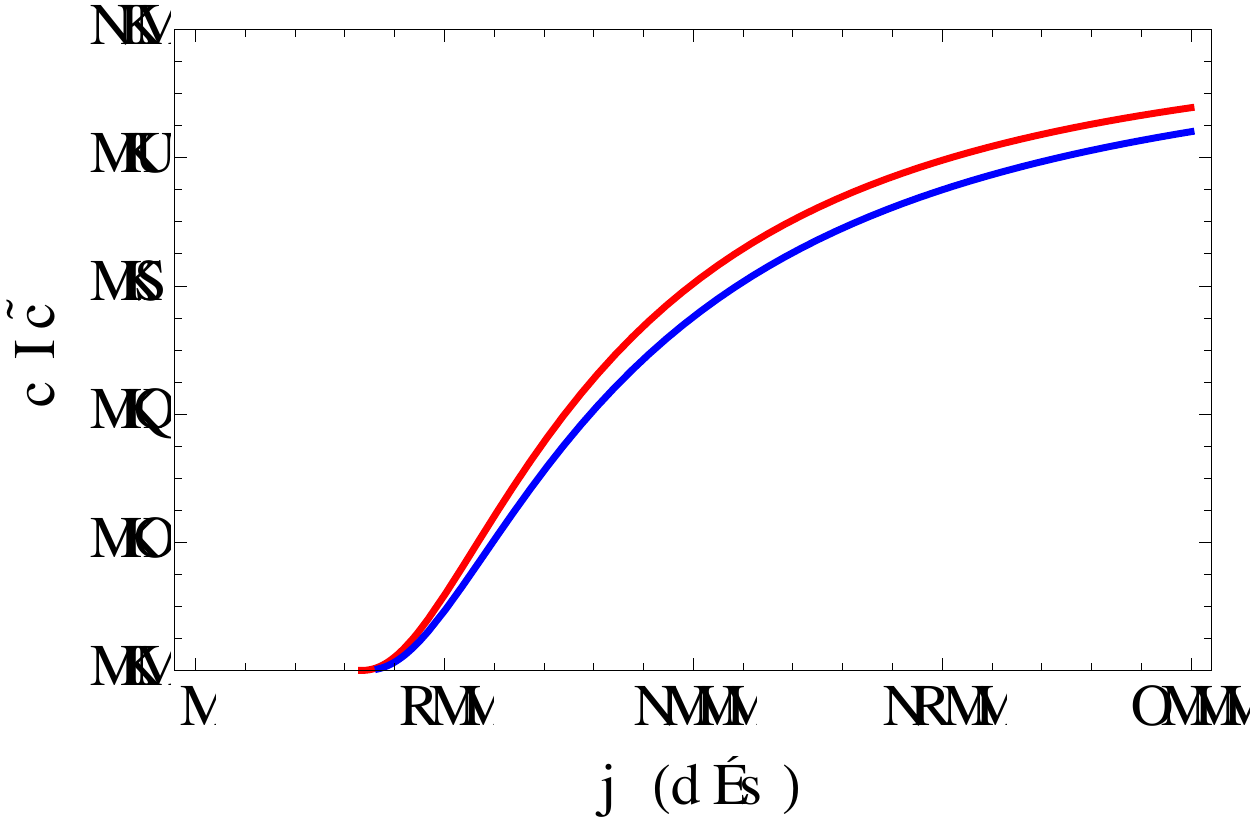}
     \includegraphics[width=0.48\textwidth]{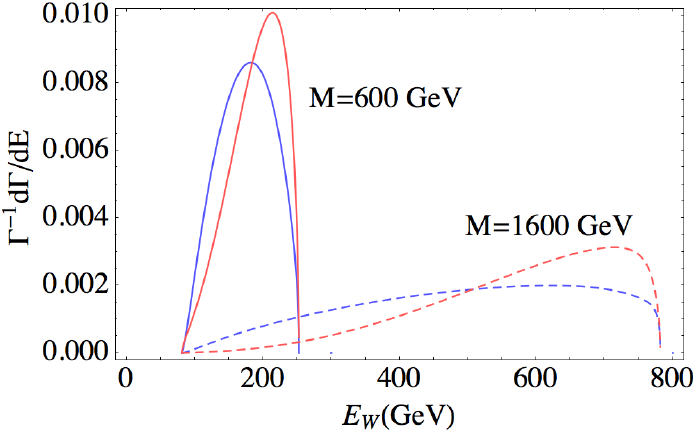}
  \end{center}
\caption{\emph{Left panel: the functions $F$ (blue) and $\tilde F$ (red) as a function of $M$: they asymptote to unity for $M\to\infty$, which is equivalent to massless final states. Right panel: the differential distribution $\Gamma^{-1}d\Gamma/dE_{W}$ for the decay $X_{8/3} \to t_L W^+W^+$ as a function of the energy of any of the two $W$s (not including the one from the top quark decay), in the $X_{8/3}$ rest frame,
 in the case of the contact interaction (blue) and the charge-5/3 mediated decay (red), for $M=600,1600\GeV$.}}\label{fig:fs}
\end{figure}

\begin{figure}[!t]
    \begin{center}
    \hspace{-0.3cm}
      \includegraphics[width=0.5\textwidth]{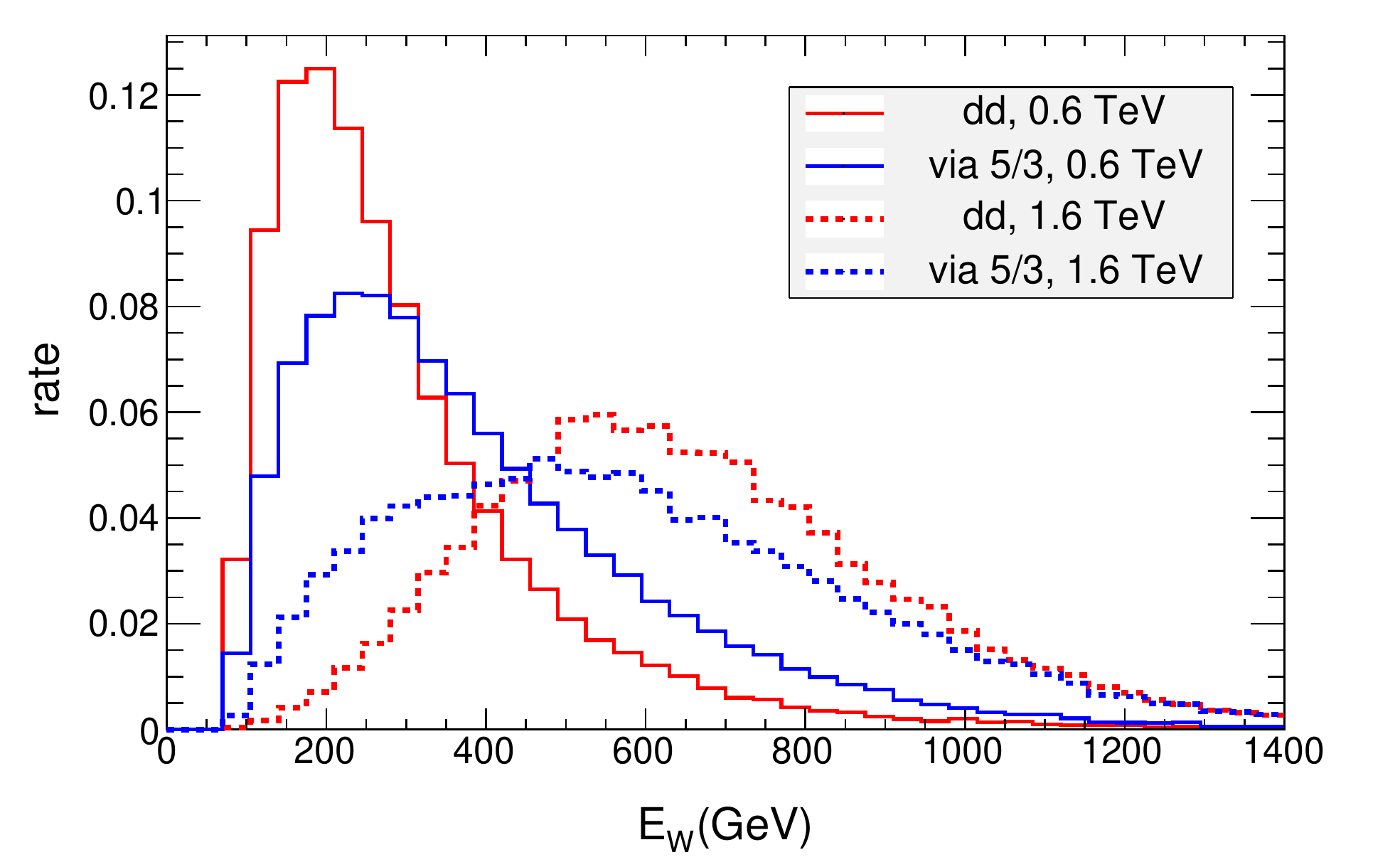}
     \includegraphics[width=0.5\textwidth]{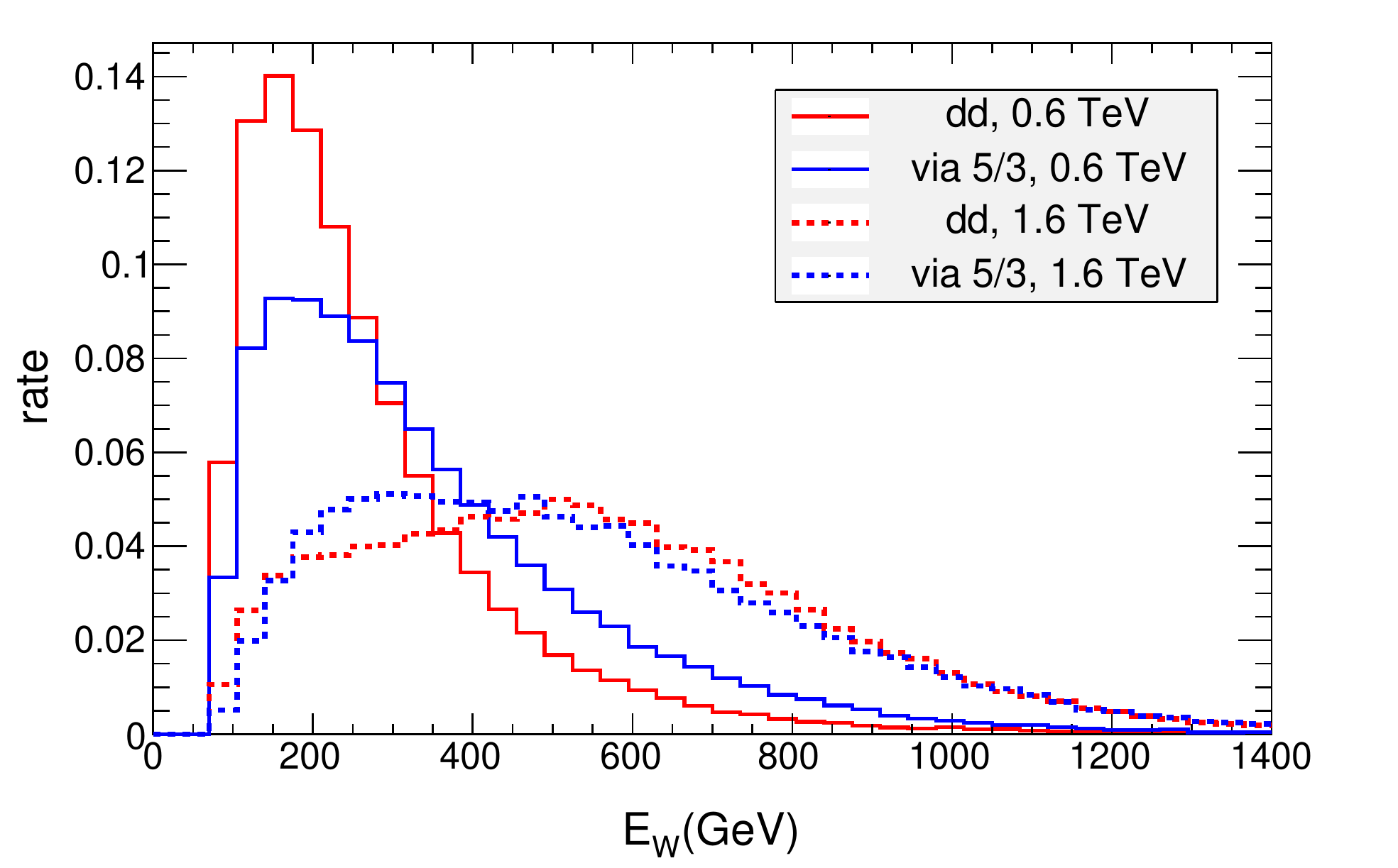}
  \end{center}
\caption{\emph{Comparison of energy distributions of $W$ bosons produced in the decays of the pair produced $X_{8/3}$ with mass 600~GeV (solid lines) and 1600~GeV (dashed lines), at LHC with 8~TeV center of mass energy, assuming that decays proceed via contact interaction (red lines) or via intermediate charge-5/3 state (blue lines). Left panel: energies of the two $W$'s from the $X_{8/3} \to W W t$ decay. Right panel: energies of all three $W$'s from the $X_{8/3} \to W W t \to WWW b$ decay.}}
\label{fig:wenergy}
\end{figure}

Hence, $\Xet$ decays only in $t W^+W^+$, and this decay can occur either via the contact interaction of \eq{dd83} or via an off-shell $\Xft$ or $\Yft$ from \eq{5383} (recall that all these states share the same mass). The contribution to the decay widths mediated by the contact interaction is
\begin{equation}
\Gamma_{dd}(\Xet\to t_L W^+W^+)=\frac{c_2^2}{7680 \pi^3}\frac{M^7}{f^4M_*^2}F\left(\frac{m_t^2}{M^2},\frac{m_W^2}{M^2}\right)
\end{equation}
while the contribution from the diagram with charge-$5/3$ states, in the limit where $\Gamma(X_{5/3})=\Gamma(Y_{5/3})=0$, is
\begin{equation}
\Gamma_{5/3}(\Xet\to t_L W^+W^+)=\frac{c_1^2 y_R^2}{768 \pi^3}\frac{v^2}{f^2}\frac{M^3}{f^2}\tilde F\left(\frac{m_t^2}{M^2},\frac{m_W^2}{M^2}\right)
\end{equation}
where $F$ and $\tilde F$ are functions of comparable shape, encoding the effects of massive final states, they are shown in the left panel of Fig.~\ref{fig:fs} and asymptote to 1 for $M\to \infty$. 
The ratio between the two widths is
\begin{equation}
\frac{\Gamma_{dd}(\Xet\to t_L W^+W^+)}{\Gamma_{5/3}(\Xet\to t_L W^+W^+)}\approx \frac{c_2^2}{c_1^2y^2}\frac{0.1}{v^2/f^2}\frac{M^4}{f^2M_*^2}
\end{equation}
and shows that  if  $M_*$ lies in the multi-TeV range (in our analysis we take $M_*=3\TeV$), the BR to contact-interaction mediated channel is typically smaller. Despite the total widths being comparable, the differential distributions of the decay products differ substantially, as shown in the right panel of Fig.~\ref{fig:fs} for the $\Xet \to WW t$ decays in the center of mass frame: the contact interaction (in red) prefers higher energy $W$'s, while the $\Xft$-mediated decay (blue) shows a slight preference for small momenta. 
The energy distributions of the decay products give an important information about how easily they would be able to pass hard $p_t$  cuts\footnote{Given that in the pair production process no preferred direction is present, the shapes of $p_t$ distributions will resemble the ones of the energy distributions.} which are typically needed to suppress the backgrounds. The behaviour of $E_W$ significantly changes if the initial $\Xet$ is boosted, which is the case for relatively low $M_{8/3}$ (Fig.~\ref{fig:wenergy}, left, solid lines), and the positions of the peaks switch places: the contact interaction now tends to produce less energetic $W$'s compared to 5/3-mediated decays. If we now also consider the third $W$, produced in the top quark decay, the overall behaviour will not change for small $M_{8/3}$ (Fig.~\ref{fig:wenergy}, right, solid lines), but at higher masses both decay modes will produce almost identical decay spectra (Fig.~\ref{fig:wenergy}, right, dashed lines). This means that for higher $\Xet$ masses, at the edge of experimental sensitivity, the signatures of both decay modes become similar, therefore one can expect that the predictions obtained with a simplified model of the Section.~\ref{sec:simplemodel}, containing only the contact interaction, will closely follow the ones of the full model.

For the 5/3-mediated decays, the differential distribution of width is peaked in the region with nearly on-shell $W$'s and an off-shell charge 5/3 state.    
Off-shell $W$'s are always disfavored:
an off-shell $W$ costs $\sim \Gamma_W^2/m_W^2$ to the total cross section, while the gain in the $\Xft$ propagator is only $\sim(1+4 m_W^2/\Gamma_{5/3}^2)$.

\subsection{Bounds from Two and Three Same-Sign Leptons Searches}
 
Given the distinctive particle content, summarized in \eq{decomposition}, one of the best search strategies for this type of models is in the \emph{2ssl} topology resulting from the decays of charge 8/3, 5/3 and -1/3 states. In fact, although dedicated searches for resonances  with charge 2/3 have reached a sensitivity comparable to searches for  charge $5/3$ states \cite{CMS23}, in this model the \emph{2ssl} topology receives contributions from a larger number of  states.

In this subsection we first recast the current $2ssl$ analyses~\cite{cms53} and the estimates for the 14 TeV LHC run~\cite{cms14} into bounds on the full model of \eq{LagCCWZ}. We sum the $2ssl$ signal from the pair-produced~\footnote{
As explained in section~\ref{sec:simplemodel}, single production is suppressed with  respect to pair production for the $\Xet$. Similarly, for members of the $X$ multiplet a suppression of $v/f$ has to be paid.
Members of $Y$ and $Z$ multiplets can be singly produced in association with $t$ or $b$, through eq.~(\ref{eq:lagrgold1}):  the case of a top quark is disfavored by its large mass, while couplings with the $b$ (for $Y_{2/3}$, $Z_{-4/3}$, $Z_{-1/3}$ and $Z_{2/3}$) can enhance single production due to the small $b$ mass. This production channel can be potentially important \cite{DeSimone:2012fs,Vignaroli}.}
 charge-$5/3$ states $\Xft,\Yft$~\footnote{Interference effects, which could in principle increase the cross-section, are suppressed by the small ratio of the $\Xft$ to the $\Yft$ widths, and have been neglected.}
 to that of $\Xet$ and include the smaller contributions from $\Yot$ and $\Zot$

\begin{equation}
N_{signal} = {\cal L} \sum_{n} \, \textrm{BR}_{n} \, \epsilon_{n} \, \sigma(M_n)\, ,
\label{eq:nsignal}
\end{equation}
where the index $n$ runs over all states with charges $8/3$, $5/3$ and $-1/3$ and  we are only including the BR that lead to \emph{2ssl} signals. Notice that  the $\Xet$ enters into the sum~(\ref{eq:nsignal})  in three different ways, weighed by  corresponding BR:  for decays via contact interactions,  for decays via charge 5/3 states, and  for a case when the first of the produced $\Xet$ decays via contact interaction, and the second via charge 5/3 (for the latter case we take the average of the efficiencies of the first two cases).

The relevant acceptances are presented in table~\ref{tab:seff1}. 
  \begin{figure}[!t]
    \begin{center}
    \hspace{-1.cm}
      \includegraphics[width=0.47\textwidth]{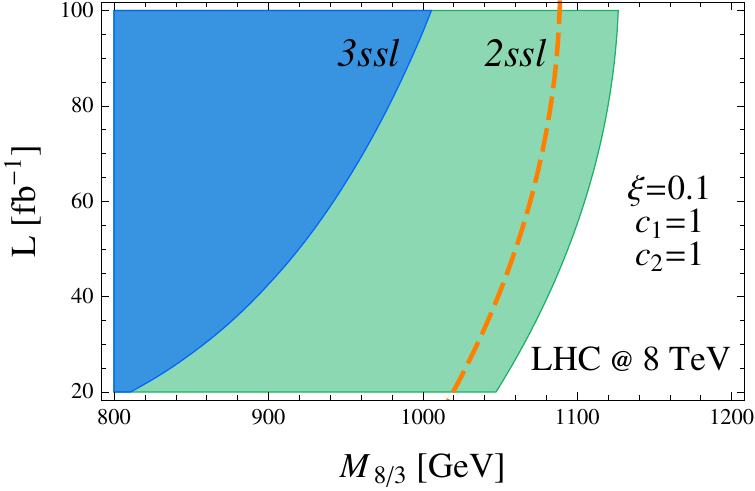}
      \includegraphics[width=0.47\textwidth]{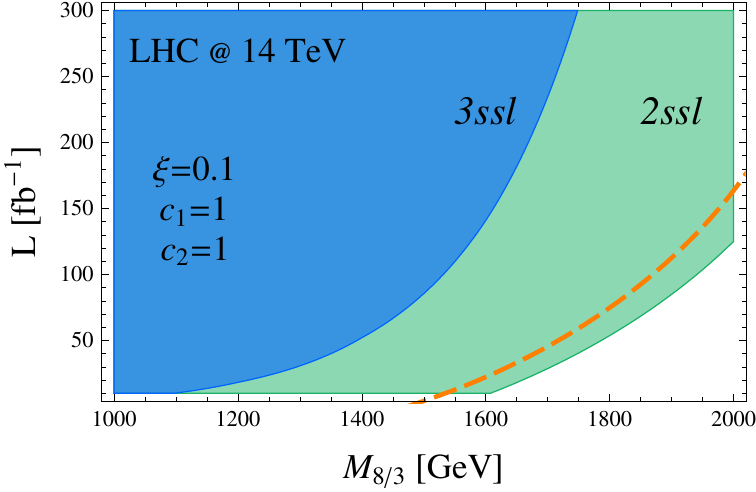}
  \end{center}
\caption{\emph{
A comparison of expected excluded masses of the charge $8/3$ state for the $2ssl$ (green) and $3ssl$ (blue) search channels in the complete model, with all the states of the nine-plet contributing to the signal, for different integrated luminosities for 8 TeV (left panel) and 14 TeV (right panel) experiments.
Orange dashed lines correspond to the exclusion provided by the 2ssl channel alone, assuming that only $X_{8/3}$ is produced.}}\label{fig:exclusion_2}
\end{figure}
The efficiency of the cuts for charge $5/3$ and $-1/3$ states are very close to each other due to the similar topology of their decays~\cite{Contino:2008hi}, but the contribution of the latter is  suppressed by the BR into $2ssl$. 
Consequently, the full signal is  mostly determined by the charge $5/3$ and $8/3$ states 
and therefore depends on the single parameter $M$ defining their masses. 
Using current data, we obtain a lower bound  
\begin{equation}
M \geq 990\,\GeV \quad @\,\,\,\, 95 \% \,\,\,\textrm{C.L.}
\end{equation}
which is marginally stronger than the bound obtained assuming that only the  $X_{8/3}$ is present (\eq{simplebound}).  Furthermore, given that for moderate values of $M$, the experimental sensitivity to  contact interactions and to the $5/3$-mediated ones are comparable, we conclude that  the constraint on  $M$ obtained in section~\ref{sec:simplemodel} for the $\Xet$ alone, is independent of the details of the model and that the bound obtained by considering $\Xet$ going to $2ssl$ alone is very close to the bound on the masses of composite resonances which can be obtained by considering the signal coming from all the states present in the model.

Finally, we perform a comparison of the sensitivities of $2ssl$ and $3ssl$ searches, using the same $3ssl$ analysis as in the Section~\ref{sec:simplemodel}, but extended for one additional decay channel via charge 5/3 states. The acceptances of the cuts for both decay channels are given in the  Table~\ref{tab:seff3ssl}. The plots in Fig.~\ref{fig:exclusion_2} show that, also in this case, the $2ssl$ channel has a better exclusion power.

\subsection{Bounds from Searches for Black Holes}

Searches for microscopic black holes~\cite{atlas:bh,Chatrchyan:2013xva},  looking for events with a large number of hard jets and leptons in the final states, are sensitive to the pair produced $\Xet$, giving rise to  6 $W$ bosons and 2 $b$-quarks, which subsequently can decay into 14 hard objects.   To understand whether such a generic signature, without any requirement on the number of leptons, can be as distinctive as the $2ssl$, we recast the search presented in  Ref.~\cite{Chatrchyan:2013xva}.
The event selection in Ref.~\cite{Chatrchyan:2013xva} is based on imposing  hard cuts on individual transverse momenta ($p_T \geq 50$ GeV), scalar sum of transverse momenta of the objects in the event ($S_T \geq 1200...5000$ GeV) and the number of constituents ($N_{con} \geq 2...10$), including jets and leptons; the bound on the BSM cross section is then obtained for every combination of $S_T$ and $N_{con}$ cuts. 

\begin{figure}[!t]
    \begin{center}
    \hspace{-1.cm}
      \includegraphics[width=0.46\textwidth]{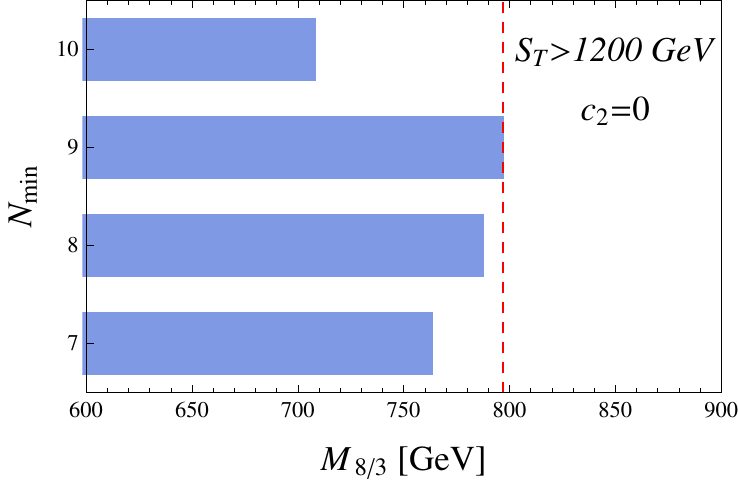}
      \includegraphics[width=0.46\textwidth]{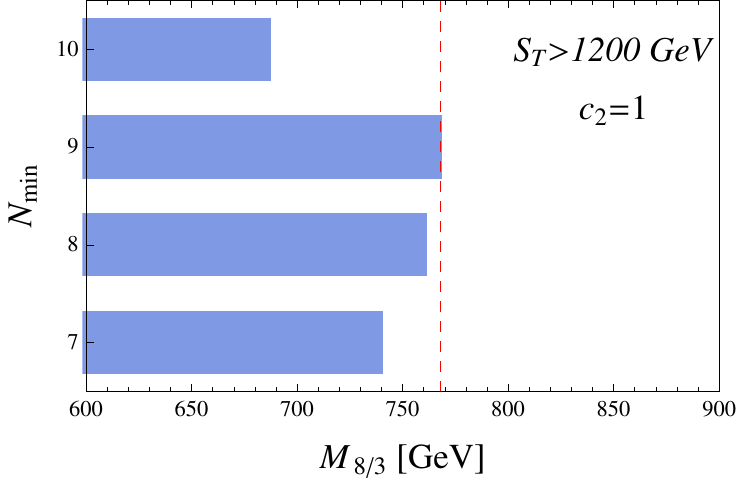}
  \end{center}
\caption{\emph{Masses of the $X_{8/3}$ excluded by the search for black holes~\cite{Chatrchyan:2013xva} in case of the decays via charge 5/3 states (left panel, $c_2=0$) and when both decay channels are present (right panel, $c_2=1$), for $\xi=0.1$ and $y=1$. The optimal exclusion ($\sim800$ GeV) is obtained for $min(S_T)$=1200 GeV and by requiring at least 9 constituents in the event.}}\label{fig:exclusion_bh}
\end{figure}

We compute the efficiency of the $\Xet$ signal for $M=600,...,1000$ GeV, $N_{con}\geq 5,...,10$ and $S_T \geq 1200, 1400, 1600$ GeV and compare the signal cross section with the bound of  Ref.~\cite{Chatrchyan:2013xva}.
It turns out that when  the $S_T$ cut is increased, the signal drops faster than the bound gets tighter: for this reason the strongest bound is obtained for the lowest value of the $S_T$ cut, 1200 GeV. 
In Fig.~\ref{fig:exclusion_bh} we show the dependence of the $M$ bound on the minimal required number of constituents in the events.

Notice that, at small mass, the efficiency of the decays via the contact interaction (and hence also for the Simplified Model of the Section~\ref{sec:simplemodel}) is relatively low  and no bound can be obtained if only this interaction is present (equivalently, for large values of $c_2$).
The situation changes however at higher $\Xet$ masses, where the efficiencies become comparable for both decay channels. One can therefore expect that, in the future, an updated analysis with improved sensitivity  will put similar bounds on the $\Xet$ of the Simplified Model and on the one of the full model, and hence the former can again be used as a good approximation to the latter.  

The sensitivity of this type of searches to composite models can be improved. This can be achieved, as already commented, by lowering the $S_T$ cut\footnote{With more statistics, however, experiments might become sensitive to values of $M$ large enough to make the signal insensitive to the $S_T$ cut.} or by using  boosted techniques in the analysis.

\section{Conclusions and Outlook}

We have studied the phenomenology of models that contain  fermionic colored states with electric charge 8/3. These are automatically present, for instance,  in composite Higgs models, whose low-energy spectrum contains composite resonances transforming as ({\bf3,3})$_{2/3}$ multiplets of  the $SU(2)_L\times SU(2)_R\times U(1)_X$ symmetry group, where the hypercharge is realized as $Y=T_R^3+X$. 
This kind of models, as  recently shown in  Refs.~\cite{Panico:2012uw,Pappadopulo:2013vca}, allows to accommodate the Higgs boson with a mass 125~GeV without a large tuning of parameters.   
Using the CCWZ construction, we have built an effective low-energy description of composite Higgs models of this type.  Its symmetry structure requires, beside the charge-8/3 state,  also the presence of resonances with electric charges $Q=5/3,2/3,-1/3$ and $-4/3$. 
We have then compared the collider signatures of this effective model with an even simpler one, in which only the charge-8/3 state is present and decays with 100\% probability to $W^+W^+t$, via a contact interaction. The latter model is {\it a priori} not related to composite Higgs scenarios and could arise from some different setup.  

From the collider phenomenology point of view, the simplified model, containing $\Xet$ only, is a very good approximation to the more complete setups, containing more states. This is because the channels allowing for an efficient separation of the background and the signal (characterised by two same sign leptons, three same sign leptons, large number of constituents) are all dominated by the decays of the state with charge $8/3$.  

With the presently collected data we found that the strongest bound $M\geq 940 \GeV$ at 95\% CL can be put on the mass of the charge 8/3 state using two same sign leptons searches if it is not accompanied by additional composite resonances, and if the other members of the ({\bf 3},{\bf 3})$_{2/3}$ are accounted for, the bound just slightly increases to $M\geq 990 \GeV$.

The $3ssl$ channel, a unique feature of the $\Xet$, appears to be less competitive than the $2ssl$, but is still important, since it provides a complementary information about the presence of  $\Xet$ in the spectrum. On the other hand, searches considering final states with a large number of constituents, tailored for the study of microscopic black holes, are at present not competitive, but could be improved.

The sensitivity of the $2ssl$ can be potentially improved by requiring additional jets or harder cuts.  Alternatively, searches for one lepton plus jets in the final state,  already shown to be more efficient than $2ssl$ in the case of $\Xft$~\cite{azatov}, have the advantage of a larger branching fraction compared to $2ssl$ due to only one leptonic $W$, while the larger background can be suppressed by requiring a large number of hard jets, which are easily produced by $\Xet$. 
Furthermore, at high masses the single production cross section can become competitive with the pair production one, due to a lower production threshold. Therefore the reach of the LHC in $\Xet$ mass exclusion can potentially be even higher than our estimates.

Our results imply that the minimally tuned model of  Ref.~\cite{Pappadopulo:2013vca}, predicting a relatively light nine-plet, with a mass lighter than 1~TeV, is already in some tension with the experimental data, while a non-observation of the $\Xet$ signal by the upcoming LHC experiments will be able to push the model into a significantly fine-tuned regime, weakining the original motivation for studying it. 

\section*{Acknowledgments}
We are particularly grateful to R.~Rattazzi for advice during the completion of this work and to A.~Wulzer for reading and commenting the manuscript. We are also glad to thank  V.~Hirschi, A.~Avetisyan, D.~Pappadopulo, G.~Perez, D.~Pirtskhalava for very useful discussions and thank the MadGraph Team for allowing us to run part of our simulations on their cluster. O.M. and F.R. acknowledge the
Galileo Galilei Institute in Florence for hospitality during the completion of this work.  F.R. also acknowledges support from the Swiss National Science Foundation, under
the Ambizione grant PZ00P2~136932 and thanks IFAE, Barcelona, for hospitality during completion of this work. The work of O.M. was supported by the European
Programme Unification in the LHC Era, contract PITN-GA-2009-237920 (UNILHC), by MIUR under the contract 2010 YJ2NYW-010 and in part by the MIUR-FIRB grant RBFR12H1MW.
We are also grateful to Heidi and grant SNF Sinergia CRSII2-141847 for provided computational resources.



\end{document}